\documentclass[pre,twocolumn,aps,showpacs]{revtex4-1}
\newcommand{\bec}[1]{\mbox{\boldmath $ #1$}}
\usepackage{bm}
\usepackage{graphicx}
\usepackage{color}

\newcommand{\meanUU}{\overline{\bm{U}}}
\newcommand{\meanWW}{\overline{\bm{W}}}
\newcommand{\meanU}{\overline{U}}
\newcommand{\meanV}{\overline{V}}
\newcommand{\meanS}{\overline{S}}
\newcommand{\meanW}{\overline{W}}
\newcommand{\meanPhi}{\overline{\Phi}}
\newcommand{\meanOmega}{\overline{\Omega}}
\newcommand{\meanP}{\overline{P}}
\newcommand{\meanT}{\overline{T}}
\newcommand{\meanrho}{\overline{\rho}}

\begin{document}

\title{Generation of a large-scale vorticity in a
fast rotating density stratified turbulence or turbulent convection}

\author{Igor Rogachevskii}
\email{gary@bgu.ac.il}
\homepage{http://www.bgu.ac.il/~gary}

\author{Nathan Kleeorin}
\email{nat@bgu.ac.il}

\affiliation{
Department of Mechanical Engineering, Ben-Gurion University of the Negev, P. O. B. 653, Beer-Sheva
 8410530, Israel
 \\
Nordita, Stockholm University and KTH Royal Institute of Technology, 10691 Stockholm, Sweden
}

\date{\today}
\begin{abstract}
We find an instability resulting in generation of large-scale vorticity
in a fast rotating small-scale turbulence
or turbulent convection
with inhomogeneous fluid density along the rotational axis in anelastic approximation.
The large-scale instability causes excitation of two modes: (i) the mode with
dominant vertical vorticity and with the mean velocity being independent of the
vertical coordinate; (ii) the mode with dominant horizontal vorticity and
with the mean momentum being independent of the vertical coordinate.
The mode with the dominant vertical vorticity can be excited in a fast rotating
density stratified hydrodynamic turbulence or turbulent convection.
For this mode, the mean entropy is depleted inside the cyclonic vortices,
while it is enhanced inside the anti-cyclonic vortices.
The mode with the dominant horizontal vorticity can be excited only in a
fast rotating density stratified turbulent convection.
The developed theory may be relevant for explanation of an origin of
large spots observed as immense storms in great planets, e.g., the Great Red Spot in Jupiter
and large spots in Saturn. It may be also useful for explanation of an
origin of high-latitude spots in rapidly rotating late-type stars.
\end{abstract}

\maketitle

\section{Introduction}

Generations of large-scale vorticity in turbulent flows have been investigated
theoretically, experimentally and numerically in a number of
studies due to various applications in geophysical, astrophysical
and industrial flows (see, e.g., \cite{L83,P87,C94}).
Using an analogy between the induction
equation for magnetic field and the vorticity equation (see \cite{B53,M78}),
it has been proposed in \cite{MST83,KMT91,CMP94} that the large-scale vorticity
can be generated due to a large-scale instability by the kinetic $\alpha$ effect
in a helical turbulence with a net kinetic helicity.
The kinetic helicity and the kinetic $\alpha$ effect can be produced in rotating
density stratified or inhomogeneous turbulence.

Another possibility for a generation of large-scale vorticity is related to
anisotropic kinetic $\alpha$ effect referred as the AKA effect \cite{FSS87,FSS88,KRK94},
which is caused by a non-Galilean invariant forcing.
For example, boundaries can break the Galilean
invariance which results in an anisotropic kinetic $\alpha$ effect \cite{BR01},
resulting in a large-scale instability.
In astrophysics, a turbulence driven by non-Galilean invariant forcing
can exist in galaxies (e.g.,
supernova-driven turbulence \citep{KKBS99,KGVS18} and
the turbulent wakes driven by galaxies moving
through the galaxy cluster \citep{RSS89}).

In a non-conducting fluid, a non-helical turbulence
with an imposed large-scale velocity shear
can cause a large-scale instability resulting in
generation of the large-scale vorticity
due to a combined effect of the large-scale shear motions and Reynolds
stress-induced production of perturbations of mean
vorticity \citep{EKR03,EGKR07}.
This effect referred as ''vorticity dynamo" has been also confirmed
in direct numerical simulations (DNS) \cite{YHSKR08,PMB09}.
This mechanism of the generation of the large-scale vorticity
is also associated with the Prandtl's first and second kinds of secondary flows
\cite{P52,B87}. In particular, the skew-induced streamwise mean vorticity generation
arises at the lateral boundaries of three-dimensional thin shear layers
and corresponds to the Prandtl's first kind of secondary flows.
In turbulent flows, the streamwise mean vorticity can be generated
by the Reynolds stress, and this mechanism is associated with formation of
the Prandtl's second kind of turbulent flows \cite{B87}.

A large-scale vorticity also can be produced by a combined effect of a rotating
incompressible turbulence and inhomogeneous kinetic helicity
\cite{YY93,YB16,IYH17,KR18}
or due to a combined action of a density stratified
rotating homogeneous turbulence and uniform kinetic helicity \cite{KR18}.
These effects result in the formation of a large-scale shear,
and in turn its interaction with the small-scale turbulence
causes an excitation of the large-scale instability (the vorticity dynamo)
due to a combined effect of the large-scale shear and Reynolds stress-induced generation
of the mean vorticity \cite{KR18}.

Recent DNS have shown that large-scale vortices in rapidly rotating turbulent convection
can be formed in compressible \cite{C07,PMH11,MPH11}
or Boussinesq fluids \cite{CHJ14,CHJ15,RJK14,FSP14,FGK19}.
The produced large-scale motions include cyclonic vortices and anti-cyclonic vortices.
It was found that in the cyclonic vortices the temperature is depleted \cite{PMH11,MPH11}.

In the present study we develop a theory of the generation of the large-scale vorticity
in a fast rotating turbulent convection
with inhomogeneous fluid density
in anelastic approximation. A particular case when gravity is along rotational axis has been considered.
We have found a large-scale instability which results in an excitation of two modes.
For the mode with dominant vertical vorticity, the mean velocity is independent
of the vertical coordinate.
This mode can be excited in both, a fast rotating density stratified hydrodynamic turbulence
and fast rotating density stratified turbulent convection.
We have demonstrated that for this mode, the mean entropy is depleted inside the cyclonic vortices
in agreement with \cite{PMH11,MPH11}.
For the second mode, the horizontal component of the mean vorticity is dominant,
and the mean momentum is independent of the vertical coordinate.

This study may be relevant to formation mechanisms of large spots observed in the form of
immense storms in great planets (e.g., the Great Red Spot in Jupiter and
large spots in Saturn, see, e.g., \cite{M93,SL91,HA07}),
and it may be useful for explanation of an origin of high-latitude spots seen in Doppler imaging in rapidly
rotating late-type stars \cite{PMH11,MPH11}.

This paper is organized as follows.
In Section II we consider the effect of fast rotation on the Reynolds stress
and the effective force.
Here we outline the method of derivations and approximations made for study of this
effect.
Using mean-field equations and the derived rotational contributions to
the Reynolds stress, we study in Section III the large-scale instability
causing the generation of the large-scale vorticity
in a fast rotating turbulent convection
with inhomogeneous fluid density along the rotational axis.
Finally, conclusions are drawn in Section V.
In Appendix A we present details  of the derivation of equation for the
rotational contributions to the Reynolds stress.
In Appendix~B we give an explicit form for the mean-field equations
describing the large-scale instability
which results in generation of the mean vorticity
for different modes.
In Appendix~C we discuss the role of the centrifugal force
in the production of large-scale vorticity.
The centrifugal force causes the inhomogeneous density distribution
in the plane perpendicular to the angular velocity ${\bm \Omega}$.
We have shown in Appendix~C that
a combined effect of a fast rotation and horizontal inhomogeneity of the fluid density
(caused by the centrifugal force)
results in the production of the large-scale vertical vorticity in an
anisotropic isothermal turbulence.

\section{Effect of fast rotation on the Reynolds stress and the effective force}

To derive mean-field equations which describe generation of the large-scale
vorticity, we consider a small-scale low-Mach-number
fast rotating density stratified turbulent convection
in anelastic approximation with equation of state for the ideal gas.
To investigate effect of fast rotation on the Reynolds stress
in a turbulent convection with inhomogeneous fluid density,
we use a mean field approach whereby the
velocity, pressure and entropy are decomposed in
the mean and fluctuating parts.
An ensemble averaging of the momentum and entropy equations yields
the equations for mean velocity, $\meanUU(t,{\bm x})$,
and mean entropy, $\meanS(t,{\bm x})$,
in the reference frame rotating with the constant angular velocity ${\bm \Omega}$:
\begin{eqnarray}
 {\partial \meanU_i \over \partial t} &+& (\meanUU \cdot
\bec{\nabla}) \meanU_i = - \nabla_i \left({\meanP \over \rho_0} \right) - g_i \meanS
+ 2 (\meanUU \times {\bm \Omega})_i
\nonumber\\
&-&  {1 \over \rho_0} \nabla_j \left(\rho_0 \langle u'_{i} \, u'_{j}\rangle \right),
\label{SSA5}\\
{\partial \meanS \over \partial t} &+& (\meanUU \cdot
\bec{\nabla}) \meanS = - (\meanUU \cdot \bec{\nabla}) S_0 - {1 \over \rho_0} \bec{\nabla} \cdot \left(\rho_0 \langle {\bm u'} \, s' \rangle \right) ,
\nonumber\\
\label{SSA6}
\end{eqnarray}
where $\meanS= \meanT/T_0 - (1-\gamma^{-1})\meanP/P_{0}$, $\meanT$ and $\meanP$ are the mean entropy, the mean temperature and the mean pressure, respectively,
$\gamma$ is the ratio of specific heats,
${\bm u}'$ and $s'$ are fluctuations of the fluid velocity and entropy,
$\langle u'_{i} \, u'_{j}\rangle$ is the Reynolds stress
describing turbulent viscosity and rotational effects to turbulent convection,
$\langle {\bm u'} \, s' \rangle$ is the turbulent flux of entropy,
$T_0$, $P_0$, $S_0$ and $\rho_0$ are the fluid temperature, pressure, entropy,
and density, respectively, in the basic reference state and
$\bec{\nabla} S_0 = (\gamma P_{0})^{-1} \bec{\nabla} P_{0}
- \rho_{0}^{-1} \bec{\nabla} \rho_{0}$.
The variables with the subscript $ "0" $ correspond to the hydrostatic
nearly isentropic basic reference state defined by
$\bec{\nabla} P_{0} = \rho_{0} {\bm g}$ and ${\bm
g} \cdot \bec{\nabla} S_0 \approx 0$,
where ${\bm g}$ is the acceleration due to the gravity.
In Eqs.~(\ref{SSA5})--(\ref{SSA6}) we neglect small
molecular viscosity and heat conductivity terms.

To derive equations for the
rotational contributions to the Reynolds stress,
we follow the method that is developed in \cite{KR18,RK18} and outlined
below (see, for details Appendix A).
We use equations for fluctuations of velocity ${\bm
u}'$ and entropy $s'= \theta/T_0 - (1-\gamma^{-1})p'/P_{0}$:
\begin{eqnarray}
{\partial {\bm u}' \over \partial t} &=& - (\meanUU \cdot
\bec{\nabla}) {\bm u}' - ({\bm u}' \cdot \bec{\nabla}) \meanUU -
\bec{\nabla} \biggl({p' \over \rho_{0}}\biggr) - {\bm g} \, s'
\nonumber\\
& & + 2 {\bm u}' \times {\bm \Omega} + {\bm U}^{N},
\label{KA1} \\
{\partial s' \over \partial t} &=& - ({\bm u}' \cdot \bec{\nabla}) \meanS
- (\meanUU \cdot \bec{\nabla}) s' + S^{N},
\label{KA2}
\end{eqnarray}
where $p'$ and $\theta$ are fluctuations of fluid
pressure and temperature, respectively,
${\bm U}^{N}=\langle ({\bm u}' \cdot \bec{\nabla}) {\bm u}'
\rangle - ({\bm u}' \cdot \bec{\nabla}) {\bm u}'$ and $S^{N}=\langle ({\bm u}' \cdot
\bec{\nabla}) s \rangle - ({\bm u}' \cdot \bec{\nabla}) s$
are the nonlinear terms,
and the angular brackets imply ensemble averaging.
In Eqs.~(\ref{KA1})--(\ref{KA2}) we neglect small
molecular viscosity and heat conductivity terms.
Equation~(\ref{KA1}) is written in the reference
frame rotating with the constant angular velocity ${\bm \Omega}$.
The turbulent convection is considered as a
small deviation from a well-mixed adiabatic
reference state.
The equations for fluctuations of
velocity and entropy are obtained by subtracting
equations~(\ref{SSA5})--(\ref{SSA6}) for the mean fields from the
corresponding equations for the total velocity $\meanUU + {\bm u}'$
and entropy $\meanS + s'$ fields.
The fluid velocity for a low Mach number
flows with strong inhomogeneity of the fluid density
$\rho_0$ along the gravity field
is assumed to be satisfied to the continuity equation written
in the anelastic approximation, ${\rm div} \,
(\rho_0 \, \meanUU) = 0$ and ${\rm div} \, (\rho_0 \, {\bm u}') = 0$.

To study the effects of fast rotation on the Reynolds stress
in density stratified turbulent convection,
we perform the derivations which include the
following steps:

(i) using new variables for fluctuations of velocity ${\bm v} = \sqrt{\rho_0}
\, {\bm u}' $ and entropy $s = \sqrt{\rho_0} \, s'$;

(ii) derivation of the equations for the
second-order moments of the velocity
fluctuations $\langle v_i \, v_j \rangle$, the
entropy fluctuations $\langle s^2 \rangle$ and
the turbulent flux of entropy $\langle v_i \, s
\rangle$ in the ${\bm k}$ space;

(iii) application of the multi-scale approach \cite{RS75}
that allows us to separate turbulent scales from large
scales;

(iv) adopting the spectral $\tau$ approximation \cite{O70,PFL76,KRR90} (see below);

(v)  solution of the derived second-order moment equations in the
${\bm k}$ space;

(vi) returning to the physical space to obtain expression for the Reynolds
stress as the function of the rotation rate $\Omega$.

The derived equations for the second-order moments of the velocity
fluctuations $\langle v_i \, v_j \rangle$, the
entropy fluctuations $\langle s^2 \rangle$ and
the turbulent flux of entropy $\langle v_i \, s
\rangle$ [see Eqs.~(\ref{KB3})--(\ref{KB5}) in Appendix~A],
include the first-order spatial differential operators
$\hat{\cal N}$  applied to the third-order moments $M^{(III)}$.
A problem arises how to close the system of the second-moment equations,
i.e., how to express the set of the third-moment terms
$\hat{\cal N} M^{(III)}({\bm k})$ through the lower moments
(see, e.g., \cite{O70,MY75,Mc90}). Various approximate methods have
been proposed to solve this problem.
In the present study we use the spectral $\tau$ approximation
(see, e.g., \cite{O70,PFL76,KRR90})), which postulates
that the deviations of the third-moment terms, $\hat{\cal N} M^{(III)}({\bm k})$,
from the contributions to these terms afforded by the background fast rotating turbulent
convection, $\hat{\cal N} M^{(III,0)}({\bm k})$, are expressed through the similar
deviations of the second-order moments, $M^{(II)}({\bm k})-M^{(II,0)}({\bm k})$
in the relaxation form:
\begin{eqnarray}
&& \hat{\cal N}M^{(III)}({\bm k}) - \hat{\cal N}M^{(III,0)}({\bm k})
\nonumber\\
&& \quad \quad \quad \quad \quad
= -{M^{(II)}({\bm k}) - M^{(II,0)}({\bm k}) \over \tau_r(k)} ,
\label{B6}
\end{eqnarray}
see for details, Eqs.~(\ref{KBBB6})-(\ref{RBBB6}) in Appendix~A.
Here the correlation functions with the superscript $(0)$
correspond to the background fast rotating turbulent
convection with zero spatial derivatives of
the mean velocity, $\nabla_{i} \meanU_{j} = 0$.
The time $\tau_r (k)$ is the characteristic relaxation time
of the statistical moments, which  can be identified with the
correlation time $\tau(k)$ of the turbulent
velocity field for large Reynolds numbers.
Validations of the $\tau$ approximation for different
situations have been performed in various direct numerical
simulations \citep{BK04,BS05,BS05B,RKKB11,HKRB12,BRK12,RKBE12,EKLR17,RKB18}
(see also discussion in Sect.~IV).

The $\tau$ approximation is a sort of the high-order closure and in general is similar to Eddy Damped
Quasi Normal Markovian (EDQNM) approximation. However some
principle difference exists between these two approaches \cite{O70,PFL76}.
The EDQNM closures do not relax to equilibrium (the background turbulence),
and the EDQNM approach does not describe properly the motions in the
equilibrium state in contrast to the $\tau$ approximation.
Within the EDQNM theory, there is no dynamically determined
relaxation time, and no slightly perturbed steady state can be
approached. In the $\tau$ approximation, the
relaxation time for small departures from equilibrium is
determined by the random motions in the  equilibrium state, but
not by the departure from the equilibrium. As follows from
the analysis in \cite{O70}, the $\tau$ approximation describes
the relaxation to the equilibrium state (the background turbulence)
much more accurately than the EDQNM approach.

We apply the $\tau$ approximation only to
study the deviations from the background turbulent convection which are
caused by the spatial derivatives of the mean velocity. The
background fast rotating turbulent convection is assumed to be known (see below).
The $\tau$ approximation is only valid for large Reynolds numbers,
where the relaxation time can be clearly identified with the turbulence correlation time.

We use the model of the background  homogeneous
turbulent convection with inhomogeneous fluid density distribution
along the gravity field which takes into account an
anisotropy of  turbulent convection caused
by the fast rotation [see Eqs.~(\ref{KB15})--(\ref{KB16})
in Appendix~A].
We assume that the background turbulent convection is of Kolmogorov type with
constant flux of energy over the spectrum,
i.e., the kinetic energy spectrum function for the
range of wave numbers $k_0<k<k_\nu$ is
$E(k) = - d \bar \tau(k) / dk$, the function $\bar \tau(k) =
(k / k_{0})^{1-q}$ with $1 < q < 3$ being the
exponent of the kinetic energy spectrum ($q =
5/3$ for a Kolmogorov spectrum).
Here $k_{\nu} = 1 / \ell_{\nu}$
is the wave number based on the viscous scale $\ell_{\nu}$,
and $k_{0} = 1 / \ell_{0} \ll k_\nu$,
where $\ell_{0}$ is the integral (energy containing) scale of turbulent motions.
The turbulent correlation time in ${\bm k}$ space
is $\tau(k) = 2\tau_{_{\Omega}} \, \bar \tau(k)$,
where the effect of rotation on the turbulent
correlation time, $\tau_{_{\Omega}}$,
is described just by an heuristic argument.
In particular, we assume that
\begin{eqnarray}
\tau_{_{\Omega}} = {\tau_0 \over [1 + C_\tau \, \Omega^2 \, \tau_0^2]^{1/2}} .
\label{AAA21}
\end{eqnarray}
Here the dimensionless constant $C_\tau \sim 1$
and $\tau_0 = \ell_{0} /u_{0}$ with the characteristic turbulent velocity
$u_0$ in the integral scale of turbulence $\ell_0$.
In particular, the squared inverse time-scale $\tau_{_{\Omega}}^{-2}$
is considered as a linear combination
of the two simple squared inverse time-scales: $\tau_0^{-2}$ and $\Omega^2$:
\begin{eqnarray}
\tau_{_{\Omega}}^{-2}= \tau_0^{-2} + C_\tau \Omega^2 .
\label{AA21}
\end{eqnarray}
For fast rotation, $\Omega \, \tau_0 \gg 1$, the
parameter $\Omega  \, \tau_{_{\Omega}}$
tends to be limiting value $C_\tau^{-1/2}$.

The above described procedure yields the rotational contribution to the Reynolds stress,
and the effective force, ${\cal F}_i^\Omega = \rho_0 \, \langle v_i \, v_j \rangle^\Omega e_j / H_\rho$
for a fast rotating density stratified turbulent convection
or for a fast rotating density stratified anisotropic homogeneous turbulence,
where $H_\rho = (|\bec{\nabla} \rho_0| / \rho_0)^{-1}$ is the density stratification hight,
$\langle v_i \, v_j \rangle^\Omega$ are
the rotational contributions to the Reynolds stress
given by Eqs.~(\ref{I1})--(\ref{I2}) in Appendix~A
and ${\bm e}$ is the vertical unit vector along the $z$ axis
(in the direction opposite to the gravity acceleration).
The components of the effective force are given by
\begin{eqnarray}
{\cal F}_x^\Omega &=& - 2 (A_F-A_u) \, \rho_0 \, \nu_{_{T}} \,
\Omega \tau_0 \, {\ell_0^2 \over H_\rho^3} \, \nabla_z \meanU_y ,
\label{I3}\\
{\cal F}_y^\Omega &=& -2 \, \rho_0 \, \nu_{_{T}} \,
\Omega \tau_0 \, {\ell_0^2 \over H_\rho^3} \, \biggl[(A_F+A_u)\nabla_x
\meanU_z
\nonumber\\
&&- (A_F-A_u) \meanW_y \biggr] ,
\label{I4}\\
{\cal F}_z^\Omega &=& -(5 A_F+ 4A_u) \, \rho_0 \,  \nu_{_{T}} \,
\Omega \tau_0 \, {\ell_0^2 \over H_\rho^3} \, \nabla_x \meanU_y ,
\label{I5}
\end{eqnarray}
where $\meanWW=\bec{\nabla} {\bm \times} \, \meanUU$ is the mean
vorticity, $\nu_{_{T}}=\tau_0 u_0^2/6$ is the turbulent viscosity,
\begin{eqnarray}
A_F &=& {9(q-1) \over 2(2q-1)} \,
{\varepsilon_{_{F}} \, \tau_0 \, F_\ast g \over \rho_0 u_0^2}  ,
\label{I6a}\\
A_u &=& {3(q-1) \over 3q-1} \, {\varepsilon_u \over 1+ \varepsilon_u} ,
\label{I6b}
\end{eqnarray}
$F_\ast = \rho_0 \, \langle u'_z  \, s' \rangle$,
the parameter $\varepsilon_u$ is the degree of anisotropy of turbulent velocity field in the background turbulence and the parameter $\varepsilon_{_{F}}$ is the degree of thermal anisotropy of the background turbulence [see Eqs.~(\ref{KB15})--(\ref{KB16}) in Appendix A].
The details of the derivation of Eqs.~(\ref{I3})--(\ref{I5}) are given in Appendix~A.
These equations are derived using the following conditions: $\Omega \tau_0 \gg 1$
and the turbulent integral scale $\ell_0$
is much smaller than the density stratification scale $H_\rho$ and the characteristic
horizontal scale $L_x$ of variations of the mean velocity $\meanUU$ (i.e., $\ell_0 \ll H_\rho; L_x$).
We also assumed that the density stratification scale $H_\rho$ is much smaller than the characteristic
vertical scale $L_z$ of variations of the mean velocity $\meanUU$.

To introduce anisotropy of turbulent velocity field in the background turbulence caused by a fast rotation, we consider an anisotropic turbulence as a combination of a three-dimensional isotropic turbulence and two-dimensional turbulence in the plane perpendicular to the rotational axis. The degree of anisotropy $\varepsilon_u$ is defined as  the ratio of turbulent kinetic energies of two-dimensional to three-dimensional motions. The degree of thermal anisotropy $\varepsilon_{_{F}}$ determines the contribution of the two-dimensional turbulence to the heat flux.

The anisotropy parameters $\varepsilon_u$ and $\varepsilon_{_{F}}$
appeared in the model of the background turbulent rotating convection
depend on the Coriolis number Co=$2 \Omega \tau_0$.
For a slow rotation (small Coriolis numbers
or large Rossby numbers), the parameters $\varepsilon_u \to 0$ and $\varepsilon_{_{F}} \to 0$.
For a fast rotation (very large Coriolis numbers or very small Rossby numbers),
the parameters $\varepsilon_u \gg 1$ and $\varepsilon_{_{F}} \sim 1$.
In this case the background turbulent convection is a highly anisotropic nearly
two-dimensional turbulence, and
the main rotational contributions to the Reynolds stress are from the two-dimensional
part of turbulence.
Formally, in the present study where we investigate a fast rotating turbulent
convection, these parameters are not specified, but they should
satisfy the following conditions $\varepsilon_u \gg 1$ and $\varepsilon_{_{F}} \sim 1$.

In the derivation of the expressions for the Reynolds stress and the effective force,
we take into account the terms which are linear in the angular velocity and
drop the terms that are quadratic in the angular velocity.
The reason is that the terms that are
proportional to the angular velocity causes generation of large-scale vorticity,
while the terms that are quadratic in the angular velocity yield small
contributions to the anisotropic part of the turbulent viscosity.
The latter effect is neglected in the present study. On the other hand,
we have taken into account the dominant contributions to the Reynolds stress
and the effective force which are caused by the effect
of fast rotation on turbulent convection.

\section{Mean-field dynamics and large-scale instability}

In this section we study large-scale instability resulting
in generation of the large-scale vorticity.
Using the derived equations~(\ref{I3})--(\ref{I5}) for the effective force,
the Navier-Stokes equation~(\ref{SSA5}) for the mean velocity $\meanUU$
and the equation for the mean vorticity $\meanWW=\bec{\nabla} {\bm \times} \, \meanUU$,
we investigate the large-scale instability.
For simplicity, we consider the case with the angular velocity
along $z$ axis (opposite to the gravity field).
The linearized equations for $\meanU_y$ and $\meanW_y$ are given by:
\begin{eqnarray}
&&{\partial \meanU_y \over \partial t}   = - 2 \, \meanU_x \Omega + {{\cal F}_y^\Omega \over \rho_0}
+ {\nu_{_{T}} \over \rho_0} \bec{\nabla} \cdot (\rho_0 \bec{\nabla} \meanU_y) ,
\nonumber\\
\label{C1} \\
&&{\partial \meanW_y \over \partial t} = 2 \Omega  \, \nabla_z \meanU_y + \left(\bec{\nabla} {\bm
\times} \, {\bec{\cal F}^\Omega \over \rho_0} \right)_y
+ {\nu_{_{T}} \over \rho_0} \bec{\nabla} \cdot (\rho_0 \bec{\nabla} \meanW_y)
\nonumber\\
&& \quad \quad - g \nabla_z \meanS .
\label{C2}
\end{eqnarray}
We introduce new variables $\meanV(t,x,z)$ and $\meanPhi(t,x,z)$:
\begin{eqnarray}
\rho_0 \meanUU = [\meanV(t,x,z) \rho_0^{1/2}] {\bm e}_y + \bec{\nabla} {\bm
\times} \, [\meanPhi(t,x,z) \rho_0^{1/2}] {\bm e}_y,
\nonumber\\
\label{CCC2}
\end{eqnarray}
which corresponds to axi-symmetric problem. In the new variables
Eqs.~(\ref{C1}) and~(\ref{C2}) are given by Eqs.~(\ref{I7}) and~(\ref{I8})
(see Appendix~B).

First, we consider a mode with the mean velocity that is independent of $z$,
i.e., we seek for a solution of Eqs.~(\ref{I7}) and~(\ref{I8}) in the following form:
$\meanV, \meanPhi \propto \exp(-\lambda z/2) \, \exp(\gamma_{\rm inst} t + i K_x X)$.
Substituting this solution into Eqs.~(\ref{I7}) and~(\ref{I8}), we obtain the growth rate
of the large-scale instability resulting in the generation of this mode:
\begin{eqnarray}
\gamma_{\rm inst} &=&\Omega \, {\ell_0^2 \over H_\rho^2} \biggl[{3 (q-1) \over 2(2q-1)}
\biggl({5 \varepsilon_{_{F}} \tau_0 \, F_\ast \, g \over \rho_0 u_0^2}
\nonumber\\
&& + {4(2q-1) \over 3(3q-1)} \, {\varepsilon_u \over 1+ \varepsilon_u}
\biggr)\biggr]^{1/2}- \nu_{_{T}} K_x^2 .
\nonumber\\
\label{I9b}
\end{eqnarray}
This mode is with a dominant vertical mean vorticity, $\meanW_z/\meanW_y \sim (H_\rho L_x) / \ell_0^2 \gg 1$,
where $L_x=2 \pi/K_x$.
It follows from Eq.~(\ref{I9b}) that the large-scale instability for this mode can be excited
even for a hydrodynamic anisotropic turbulence
(i.e., when there is no turbulent convection, $F_\ast =0$).
The mechanism of the large-scale instability resulting in the generation of the
dominant vertical mean vorticity, $\meanW_z=\nabla_x \meanU_y$,
is as follows. The Coriolis force for a fast rotation strongly modifies turbulence
and the Reynolds stress, so that the second term in Eq.~(\ref{C2}) does not vanish, $[\bec{\nabla} {\bm
\times} \, (\bec{\cal F}^\Omega / \rho_0)]_y \not=0$. This term depends on $\meanU_y$ [see Eqs.~(\ref{I5}) and~(\ref{C2})].
The horizontal component of the mean
vorticity $\meanW_y$ is produced by this key term,
$[\bec{\nabla} {\bm \times} \, (\bec{\cal F}^\Omega / \rho_0)]_y$, which is caused by the effective force,
i.e., $\partial \meanW_y / \partial t \sim [\bec{\nabla} {\bm
\times} \, (\bec{\cal F}^\Omega / \rho_0)]_y$, see Eq.~(\ref{C2}).
On the other hand, the velocity component $\meanU_y$ is produced by the Coriolis force,
$\partial \meanU_y / \partial t \sim - 2 \, \meanU_x \Omega$ [see Eq.~(\ref{C1})],
which closes the generation loop. Here we took into account that the Coriolis force is much larger than the effective force, i.e., the ratio $|2 \, \meanU_x \Omega|/
|{\cal F}_y^\Omega / \rho_0| \sim L_x H_\rho^3 /\ell_0^4~\gg~1$.

Usually for a fast rotation, inertial waves characterised by the dispersion relation,
$\omega=2 ({\bm \Omega} \cdot {\bm K})/K$, are dominant and they
decrease the growth rate of instabilities for different modes.
However, since for the considered mode the vertical derivative $\nabla_z \meanU_y=0$,
the contribution of this effect (caused by the inertial waves) to the growth rate
of the large-scale instability for this mode vanishes.

Let us study the evolution of the mean entropy $\meanS$ in this mode.
The linearised equation~(\ref{SSA6}) for $\meanS$ reads:
\begin{eqnarray}
&&{\partial \meanS \over \partial t} = - \meanU_z \nabla_z S_0 + \rho_0^{-1} \bec{\nabla} \cdot (\rho_0 \kappa_{_{T}} \bec{\nabla} \meanS) ,
\label{MM1}
\end{eqnarray}
where $\kappa_{_{T}}$ is the coefficient of turbulent diffusion.
This implies that
\begin{eqnarray}
\meanS = {\meanU_z |\nabla_z S_0| \over \gamma_{\rm inst} +
\kappa_{_{T}} K_x^2} = - {\meanW_z H_\rho \over 2 \Omega} \, \left({\gamma_{\rm inst} +
\nu_{_{T}} K_x^2 \over \gamma_{\rm inst} + \kappa_{_{T}} K_x^2} \right) \, |\nabla_z S_0|,
\nonumber\\
\label{MM2}
\end{eqnarray}
where we use the solutions for the vertical mean velocity
$\meanU_z = K_x \Phi_\ast \cos(K_x X + \varphi)
\exp(\gamma_{\rm inst} t)$, and the vertical mean vorticity
$\meanW_z = K_x V_\ast \cos(K_x X + \varphi) \exp(\gamma_{\rm inst} t)$.
Here the ratio of amplitudes $V_\ast/\Phi_\ast$ for this mode is
\begin{eqnarray}
{V_\ast \over \Phi_\ast} = - {2 \Omega \over H_\rho (\gamma_{\rm inst} + \kappa_{_{T}} K_x^2)} .
\label{MM3}
\end{eqnarray}
In Eq.~(\ref{MM3}) we neglect the small terms $\sim$ O$(\ell_0^2/H_\rho^2)$.
Thus, the solution for the mean entropy is
$\meanS = - S_\ast \cos(K_x X + \varphi) \exp(\gamma_{\rm inst} t)$.
Equation~(\ref{MM2}) implies that inside the cyclonic vortices where the perturbations of the vertical
mean vorticity are positive ($\meanW_z > 0$), the perturbations of the
mean entropy are negative ($\meanS < 0$).
Therefore, inside the cyclonic vortices the mean entropy is reduced.
On the other hand, inside the anti-cyclonic vortices where the perturbations of the vertical
mean vorticity are negative ($\meanW_z < 0$), the perturbations of the
mean entropy are positive ($\meanS > 0$).
Therefore, inside the anti-cyclonic vortices the mean entropy is increased.

There is also another mode for which the negative contribution caused by
the inertial waves to the growth rate of the instability for this mode vanishes.
Indeed, for this mode a solution of Eqs.~(\ref{I7}) and~(\ref{I8})
has the following form:
$\meanV, \meanPhi \propto \exp(\lambda z/2) \, \exp(\gamma_{\rm inst} t + i K_x X)$.
This is a mode with the mean momentum, $\rho_0 \meanUU$, that is independent of $z$.
Substituting this solution into Eqs.~(\ref{I7}) and~(\ref{I8}), we obtain the growth rate
of the large-scale instability resulting in the generation of this mode:
\begin{eqnarray}
&&\gamma_{\rm inst}= \Omega \, {\ell_0^2 \over H_\rho^2} \left[{6 (q-1) \, \varepsilon_{_{F}}
\tau_0 \, F_\ast \, g \over (2q-1) \,\rho_0 u_0^2} \right]^{1/2} - \nu_{_{T}} K_x^2 .
\label{I9}
\end{eqnarray}
This mode is with a dominant horizontal mean vorticity,
i.e., $\meanW_z/\meanW_y \sim \ell_0^2 / (H_\rho L_x) \ll 1$.
It follows from Eq.~(\ref{I9}) that the large-scale instability for this mode can be excited only in
turbulent convection (when $F_\ast \not=0)$.
For this mode the component of the mean velocity $\meanU_x=0$, and the component
$\meanU_y$ is produced by the effective force ${\cal F}_y^\Omega / \rho_0$
[see Eq.~(\ref{C1})]. On the other hand, the dominant horizontal
mean vorticity $\meanW_y$ is produced by
the term $2 \Omega  \, \nabla_z \meanU_y$
caused by the Coriolis force [see Eq.~(\ref{C2})],
which closes the generation loop.

Let us check if the obtained results are consistent with the Taylor-Proudman theorem.
For a fast rotating laminar flow, the Taylor-Proudman theorem implies that
the leading-order balance in the equation for the vorticity
for large Coriolis number (small Rossby numbers)
is $({\bm \Omega} \cdot {\bm \nabla}) {\bm U} =0$.
This implies that the velocity is independent
of the vertical coordinate $z$, where ${\bm \Omega}=\Omega {\bm e}_z$.
For the  mode with the dominant vertical mean vorticity,
the mean velocity is independent of $z$.
This implies that this mode is consistent with the Taylor-Proudman theorem.
On the other hand, for the mode with the dominant horizontal mean vorticity,
the mean momentum is independent of $z$, while the mean velocity depends on $z$,
so that this mode is not consistent with the Taylor-Proudman theorem.

\section{Discussion and conclusions}

In the present study we have considered a fast rotating turbulence or turbulent convection
with inhomogeneous fluid density along the rotational axis in anelastic approximation.
A large-scale instability exciting at large Coriolis number has been found,
which causes generation of large-scale vorticity for two key modes
with dominant vertical or horizontal components.
The effective force caused by the rotational contribution to the Reynolds stress in small-scale turbulent convection in combination with the Coriolis force in the mean-field momentum equation
are the main effects resulting in the generation of the large-scale vorticity
due to the excitation of the large-scale instability.
The mode with the vertical vorticity can be generated in both, a fast rotating
density stratified hydrodynamic turbulence and turbulent convection,
while the mode with the dominant horizontal vorticity can be excited only in a
fast rotating density stratified turbulent convection.
When the density stratification hight $H_\rho \to \infty$
(i.e., when the fluid density is uniform), the large-scale instability
found in the present study cannot be excited [see Eqs.~(\ref{I9b}) and~(\ref{I9})
for the growth rates of the instability].
This implies that this theory cannot describe formation of large-scale vortices
observed in the Boussinesq turbulent convection with ${\rm div} \, {\bm u} = 0$
(see Refs.~\cite{CHJ14,CHJ15,RJK14,FSP14,FGK19}).

Our theory is developed for a low-Mach number fast rotating turbulent convection
with inhomogeneous fluid density, which
corresponds to the set-ups of DNS described in Refs.~\cite{PMH11,MPH11}.
However, in DNS on the large-scale vorticity growth,
it is very difficult to observe the kinematic stage of the evolution
of the large-scale vorticity with an exponential growth. Usually in DNS it is only
seen the nonlinear evolution of the large-scale vorticity.
This implies that it is very difficult to make quantitative
comparisons between the kinematic mean-field theory
for the large-scale vorticity growth and DNS.
We have only performed a qualitative comparison with the DNS
described in Refs.~\cite{PMH11,MPH11}.
In particular, we confirm the existence of the threshold in the Coriolis number
for the generation of the large-scale vorticity.
The critical Coriolis number should be much larger than 1.
The derived mean-field equations describe formations
of both, cyclonic and anti-cyclonic large-scale vortices
in the kinematic (linear) stage of the instability.
As in the DNS, we also find the similar behaviour of the mean entropy or temperature
inside cyclonic and anti-cyclonic vortices.
For example, we have shown that for the mode with the dominant vertical mean vorticity,
the mean entropy is decreased inside the cyclonic vortices and
increased inside the anti-cyclonic vortices in agreement with \cite{PMH11,MPH11}.

To derive equations for the rotational contribution to the Reynolds stress
and the effective force in fast rotating density stratified turbulent convection,
we apply the spectral $\tau$ approximation (see Sect.~II).
The $\tau$ approximation is an universal tool in turbulent transport
that allows to obtain closed results and compare them with the results
of laboratory experiments, observations and numerical simulations.
The $\tau$ approximation reproduces many well-known phenomena found
by other methods in turbulent transport of particles, temperature and
magnetic fields, in turbulent convection and stably stratified
turbulent flows (see below).

In turbulent transport, the $\tau$ approximation yields correct
formulae for turbulent diffusion, turbulent thermal diffusion
and turbulent barodiffusion \cite{EKR96,BF03}.
The phenomenon of turbulent thermal diffusion
(a nondiffusive streaming of particles in the direction of
the mean heat flux), has been predicted using the stochastic calculus (the path
integral approach), the quasi-linear approach and the $\tau$ approximation.
This phenomenon has been already detected in laboratory experiments
in oscillating grids turbulence \cite{EEKR04} and in a multi-fan
produced turbulence \cite{EEKR06} in both, stably and unstably
stratified fluid flows.
The phenomenon of turbulent thermal diffusion
has been also detected in direct numerical simulations \cite{HKRB12,BRK12,RKB18}.
The numerical and experimental results are
in a good agreement with the theoretical studies performed by means
of different approaches (see \cite{EKR96,PM02}).

The $\tau$ approximation reproduces the
well-known $k^{-7/3}$-spectrum of anisotropic velocity fluctuations
in a sheared turbulence (see \cite{EKRZ02}).
This spectrum was previously found in analytical, numerical,
laboratory studies and was observed in the atmospheric
turbulence (see, e.g., \cite{L67}).
In the turbulent boundary layer problems, the
$\tau$ approximation yields correct expressions for turbulent viscosity,
turbulent thermal conductivity and the classical heat flux.
This approach also describes the counter wind heat flux
and the Deardorff's heat flux in convective boundary layers (see \cite{EKRZ02}).
These phenomena have been previously studied using different approaches
(see, e.g.,  \cite{MY75,Mc90,Z91}).

The theory of turbulent convection \cite{EKRZ02} based on the $\tau$ approximation
explains the recently discovered hysteresis phenomenon in laboratory
turbulent convection \cite{EEKRM06}. The results
obtained using the $\tau$ approximation allow also to explain
the most pronounced features of typical semi-organized coherent structures
observed in the atmospheric convective boundary layers
("cloud cells" and "cloud streets") \cite{ET85}.
The theory \cite{EKRZ02} based on the $\tau$ approximation predicts realistic values
of the following parameters: the aspect ratios of structures, the ratios of the
minimum size of the semi-organized structures to the maximum scale
of turbulent motions and  the characteristic lifetime of the
semi-organized structures. The theory \cite{EKRZ02} also predicts excitation of
convective-shear waves propagating perpendicular to the convective
rolls ("cloud streets"). These waves have been observed in the atmospheric
convective boundary layers with cloud streets \cite{ET85}.
A theory \cite{ZEKR07,ZKR08,ZKR09,ZKR13,KRZ19} for stably stratified atmospheric turbulent
flows based on both, the budget equations for the key
second moments, turbulent kinetic and potential energies and vertical turbulent
fluxes of momentum and buoyancy, and the $\tau$ approximation is in a good agrement with data from atmospheric
and laboratory experiments, direct numerical simulations and large-eddy simulations
(see detailed comparison in \cite{ZEKR07,ZKR13}).

The detailed verification of the $\tau$ approximation in the direct numerical
simulations of turbulent transport of passive scalar has been
performed in \cite{BK04}. In particular, the results
on turbulent transport of passive scalar obtained using direct
numerical simulations of homogeneous isotropic turbulence have been
compared with that obtained using a closure model based on the
$\tau$ approximation. The numerical and analytical results are in a
good agreement.

In magnetohydrodynamics, the $\tau$ approximation reproduces many
well-known phenomena found by different methods, e.g., the
$\tau$ approximation yields correct formulae
for the $\alpha$-effect \cite{KR80,RK93,RK00,RKR03}, the turbulent
diamagnetic and paramagnetic velocities \cite{Z57,VK83,RKR03},
the turbulent magnetic diffusion \cite{KR80,VK83,RKR03,RK04},
the ${\bf \Omega} {\bf \times} {\bf J}$ effect and
the $\kappa$-effect \cite{KR80,RKR03}.

The developed theory in the present study may be important for interpretation
of origin of large spots in the great planets (e.g.,
the Great Red Spot in Jupiter \cite{M93} and large spots in Saturn \cite{SL91}).
The giant planets Jupiter and Saturn have outer convection zones
of rapidly rotating convection \cite{HA07}.
The spots on giant planets are not of magnetic origin
and may be related to the large-scale instability
excited the convective turbulence.
The developed theory  may be also useful for explanation of an origin of high-latitude spots
in rapidly rotating late-type stars \cite{PMH11,MPH11}.

We have also discuss a role of the centrifugal force in
production of large-scale vorticity by a fast rotating
homogeneous anisotropic turbulence in a special case when the gravity
force is small
(see Appendix~C).
In this case the centrifugal force should be taken into account,
which causes an inhomogeneous fluid density distribution in the plane perpendicular to
the angular velocity. As a result, the large-scale vertical vorticity
is produced by a combined effect of a fast rotation and horizontal
inhomogeneity of the fluid density.

\begin{acknowledgements}
We have benefited from stimulating discussions with Axel Brandenburg, Maarit K\"apyl\"a,
Petri K\"apyl\"a and Nishant Singh.
This work was supported in part by the Israel Science Foundation governed by the Israeli
Academy of Sciences (grant No. 1210/15).
\end{acknowledgements}

\appendix

\section{Derivation of equation for the
rotational contributions to the Reynolds stress}

In this Appendix we derive equation for the
rotational contributions to the Reynolds stress.
We follow the approach developed in \cite{KR18,RK18}.
Fluctuations of velocity ${\bm u}'$ and entropy $s'$
are given by Eqs.~(\ref{KA1}) and~(\ref{KA2}).
We rewrite these equations in the ${\bm k}$ space using new variables for
fluctuations of velocity ${\bm v}= \sqrt{\rho_0}
\, {\bm u}'$ and entropy $s = \sqrt{\rho_0} \,
s'$, and derive equations for the following
correlation functions:
$f_{ij}({\bm k},{\bm K}) = \langle v_i(t,{\bm k}_1) v_j(t,{\bm
k}_2) \rangle$,
$F_{i}({\bm k},{\bm K}) = \langle s(t,{\bm k}_1) v_i(t,{\bm
k}_2) \rangle$
and
$\Theta_{i}({\bm k},{\bm K}) = \langle s(t,{\bm k}_1) s(t,{\bm
k}_2) \rangle$.
Here we apply multi-scale approach \cite{RS75}, where
${\bm k}_1 = {\bm k} + {\bm K} / 2$, $\,
{\bm k}_2 = -{\bm k} + {\bm K} / 2$, the wave vector ${\bm
K}$ and the vector ${\bm R}= ({\bm x}+{\bm y})/2$
correspond to the large scales,  while ${\bm
k}$ and ${\bm r}= {\bm y}-{\bm x}$ correspond to the  small ones. Hereafter we omitted
argument $t$ in the correlation functions. The
equations for these correlation functions are
given by
\begin{eqnarray}
{\partial f_{ij}({\bm k},{\bm K}) \over \partial t} &=& (I_{ijmn}^U +
L_{ijmn}^{\Omega}) f_{mn} + M_{ij}^F + \hat{\cal N} \tilde f_{ij} ,
\nonumber\\
\label{KB3} \\
{\partial F_{i}({\bm k},{\bm K}) \over \partial t} &=& (J_{im}^U +
D_{im}^{\Omega}) F_{m} + g e_m P_{im}({\bm k}_1) \Theta
\nonumber\\
&& + \hat{\cal N} \tilde F_{i} ,
\label{KB4} \\
{\partial \Theta({\bm k},{\bm K}) \over \partial
t} &=&  - {\rm div} \, [\meanUU \, \Theta] +
\hat{\cal N} \Theta ,
\label{KB5}
\end{eqnarray}
\\
\\
where $D_{ij}^{\Omega}({\bm k}) = 2 \varepsilon_{ijm} \Omega_n k_{mn}$,
$L_{ijmn}^{\Omega} = D_{im}^{\Omega}({\bm k}_1) \, \delta_{jn} +
D_{jn}^{\Omega}({\bm k}_2) \, \delta_{im}$,
$\delta_{ij}$ is the Kronecker unit tensor, $k_{ij} = k_i  k_j /
k^2$, $\varepsilon_{ijk}$ is the Levi-Civita fully antisymmetric tensor,
${\bm e}$ is the unit vector directed opposite
to the acceleration due to the gravity,
\begin{widetext}
\begin{eqnarray}
&& I_{ijmn}^U = J^U_{im}({\bm k}_1) \, \delta_{jn} + J^U_{jn}({\bm
k}_2) \, \delta_{im}
= \biggl[2 k_{iq} \delta_{mp} \delta_{jn} + 2 k_{jq} \delta_{im}
\delta_{pn} - \delta_{im} \delta_{jq} \delta_{np}
- \delta_{iq} \delta_{jn} \delta_{mp} + \delta_{im}
\delta_{jn} k_{q} {\partial \over \partial k_{p}} \biggr] \nabla_{p}
\meanU_{q}
\nonumber\\
&& \quad \quad \quad - \delta_{im} \delta_{jn} \, [{\rm div} \, \meanUU + \meanUU
{\bm \cdot} \bec{\nabla}],
\label{EC1}
\end{eqnarray}
\end{widetext}
\noindent
and
\begin{eqnarray}
&& M_{ij}^F = g e_m [P_{im}({\bm k}_1) F_{j}({\bm
k},{\bm K})  + P_{jm}({\bm k}_2) F_{i}(-{\bm
k},{\bm K})],
\nonumber\\
\label{EC3}\\
&& J_{ij}^U({\bm k}) = 2 k_{in} \nabla_{j} U_{n} - \nabla_{j} U_{i} -
\delta_{ij} [(1/2) \, {\rm div} \, {\bm U}
+ i ({\bm U}{\bm \cdot} {\bm k})],
\nonumber\\
\label{EC4}
\end{eqnarray}
$P_{ij}({\bm k}) = \delta_{ij} - k_{ij}$ and
$F_{i}(-{\bm k},{\bm K}) = \langle s({\bm k}_2) v_i({\bm
k}_1) \rangle$. Note that the correlation functions $f_{ij}$, $\,
F_{i}$ and $\Theta$ are proportional to the fluid density
$\rho_0({\bm R})$. Here the third-order moments
appearing due to the nonlinear terms, $\hat{\cal N}\tilde f_{ij}$, $\, \hat{\cal
N}\tilde F_{i}$ and $\hat{\cal N}\Theta$, are given by
\begin{eqnarray}
\hat{\cal N}\tilde f_{ij} &=& \langle P_{im}({\bm k}_1)
v^{N}_{m}({\bm k}_1) v_j({\bm k}_2) \rangle
\nonumber\\
&& + \langle v_i({\bm k}_1) P_{jm}({\bm k}_2) v^{N}_{m}({\bm k}_2)
\rangle ,
\label{EC5}\\
\hat{\cal N}\tilde F_{i} &=& \langle s^{N}({\bm k}_1) u_j({\bm k}_2)
\rangle + \langle s({\bm k}_1) P_{im}({\bm k}_2) v^{N}_{m}({\bm
k}_2) \rangle ,
\nonumber\\
\label{EC6}\\
\hat{\cal N}\Theta &=& \langle s^{N}({\bm k}_1) s({\bm k}_2) \rangle
+ \langle s({\bm k}_1) s^{N}({\bm k}_2) \rangle ,
\label{EC7}
\end{eqnarray}
where ${\bm v}^{N}({\bm k})$ and $s^{N}({\bm k})$ are the nonlinear terms
related to ${\bm U}^{N}$ and $S^{N}$ and rewritten in new variables.

In tensors $D_{ij}^{\Omega}$ and
$L_{ijmn}^{\Omega}$  we extract the parts which
depend on the density stratification effects,
characterised by the vector $\bec{\lambda} = -(\bec{\nabla} \rho_0) / \rho_0$,
i.e.,
\begin{eqnarray}
D_{ij}^{\Omega} &=& \tilde D_{ij} + D_{ij}^\lambda + D_{ij}^{\lambda^2} + O(\lambda^3) ,
\label{KB18}\\
L_{ijmn}^{\Omega} &=& \tilde L_{ijmn}+
L_{ijmn}^\lambda + L_{ijmn}^{\lambda^2}+ O(\lambda^3) ,
\label{KB19}
\end{eqnarray}
where $\tilde D_{ij} = 2 \varepsilon_{ijp} \Omega_q k_{pq}$,
$D_{ij}^\lambda = 2 \varepsilon_{ijp} \Omega_q k_{pq}^\lambda$,
$D_{ij}^{\lambda^2} = 2 \varepsilon_{ijp} \Omega_q k_{pq}^{\lambda^2}$,
\begin{eqnarray}
\tilde L_{ijmn} &=& 2 \, \Omega_q \,
(\varepsilon_{imp}  \, \delta_{jn} +
\varepsilon_{jnp} \, \delta_{im}) \, k_{pq},
\label{EC8}\\
L_{ijmn}^\lambda &=& - 2\,\Omega_q \,
\Big[(\varepsilon_{imp}  \, \delta_{jn} -
\varepsilon_{jnp} \, \delta_{im}) \,
k_{pq}^\lambda
\nonumber\\
&& +{i \over k^2} (\varepsilon_{ilq} \,
\delta_{jn}   \,\lambda_m- \varepsilon_{jlq} \,
\delta_{im}\,\lambda_n) \, k_{l} \Big],
\label{EC9}\\
L_{ijmn}^{\lambda^2} &=& 2\, \Omega_q \,
(\varepsilon_{imp}  \, \delta_{jn} +
\varepsilon_{jnp} \, \delta_{im}) \,
k_{pq}^{\lambda^2},
\label{EC10}\\
k_{ij}^\lambda &=& {i \over 2 k^2} \, [k_i \lambda_j
+ k_j \lambda_i - 2 k_{ij} ({\bm k} {\bm \cdot} \bec{\lambda})],
\label{E11}\\
k_{ij}^{\lambda^2} &=& {1 \over 4 k^2} \,
\big[\lambda_i \lambda_j -
k_{ij} \lambda^2  + 4 k_{ijpq}\lambda_p
\lambda_q\big] .
\label{EC12}
\end{eqnarray}

Next, we apply the spectral $\tau$ approximation [see Eq.~(\ref{B6})], i.e.,
\begin{eqnarray}
\hat{\cal N}f_{ij}({\bm k}) - \hat{\cal N}f_{ij}^{(0)}({\bm k}) = -
{f_{ij}({\bm k}) - f_{ij}^{(0)}({\bm k}) \over \tau(k)},
\label{KBBB6}\\
\hat{\cal N}F_{i}({\bm k}) - \hat{\cal N}F_{i}^{(0)}({\bm k}) = -
{F_{i}({\bm k}) - F_{i}^{(0)}({\bm k}) \over \tau(k)},
\label{SBBB6}\\
\hat{\cal N}\Theta({\bm k}) - \hat{\cal N}\Theta^{(0)}({\bm k}) = -
{\Theta({\bm k}) - \Theta^{(0)}({\bm k}) \over \tau(k)},
\label{RBBB6}
\end{eqnarray}
where $\hat{\cal N}f_{ij} = \hat{\cal N}\tilde f_{ij} +
M_{ij}^F(F^{\Omega=0})$ and $\hat{\cal N} F_{i} =
\hat{\cal N}\tilde F_{i} + g e_n P_{in}(k)
\Theta^{\Omega=0}$. The
quantities $F^{\Omega=0}$ and $\Theta^{\Omega=0}$
are for a nonrotating turbulent convection with
nonzero spatial derivatives of the mean velocity. The superscript $(0)$
corresponds to the rotating background turbulent
convection with $\nabla_{i} \meanU_{j} = 0$.

Equations~(\ref{KB3})-(\ref{KB5}) in a
steady state read
\begin{eqnarray}
f_{ij}({\bm k}) &=& L_{ijmn}^{-1}
\big[f^{(0)}_{mn} + \tau \,  \tilde M_{mn}^F +
\tau \, (I_{mnpq}^U + L_{mnpq}^{\lambda}
\nonumber\\
&& + L_{mnpq}^{\lambda^2})  \, f_{pq} \big],
\label{KB7}\\
F_i({\bm k}) &=& D_{im}^{-1}
\big[F^{(0)}_{m}({\bm k})  + \tau \, (J_{mn}^U + D_{mn}^{\lambda}
+ D_{mn}^{\lambda^2}) F_{n} \big],
\nonumber\\
\label{KB8}
\end{eqnarray}
where
\begin{eqnarray}
\tilde M_{ij}^F &=& g e_m \big\{\big[P_{im}({\bm
k}) + k_{im}^\lambda + k_{im}^{\lambda^2} \big]
\tilde F_{j}({\bm k})+ \big[P_{jm}({\bm k})
\nonumber\\
&& - k_{jm}^\lambda+ k_{jm}^{\lambda^2} \big] \tilde F_{i}(-{\bm k})
\big\},
\label{KBB8}
\end{eqnarray}
$\tilde F_{i}=F_{i}-F_{i}^{\Omega=0}$ and we
neglected small terms $\sim O(\lambda^3)$, see \cite{RK18}.
In Eqs.~(\ref{KB7})--(\ref{KB8}), the operator $D_{ij}^{-1}$ is the inverse of
$\delta_{ij} - \tau \tilde D_{ij}$ and the
operator $L_{ijmn}^{-1}({\bm \Omega})$ is the
inverse of $\delta_{im} \delta_{jn} - \tau \,
\tilde L_{ijmn}$, where
\begin{eqnarray}
&& D_{ij}^{-1} = \chi(\psi) \, (\delta_{ij} +
\psi \, \varepsilon_{ijm} \, \hat k_m + \psi^2 \, k_{ij}),
\label{KB12}
\end{eqnarray}
and
\begin{widetext}
\begin{eqnarray}
L_{ijmn}^{-1}({\bm \Omega}) &=&  {1 \over 2} \Big[B_1
\, \delta_{im} \delta_{jn} + B_2 \, k_{ijmn} +
B_3 \, (\varepsilon_{imp} \delta_{jn}
+ \varepsilon_{jnp} \delta_{im}) \hat k_p +
B_4 \, (\delta_{im} k_{jn} + \delta_{jn} k_{im})
\nonumber\\
&& + B_5 \, \varepsilon_{ipm} \varepsilon_{jqn}
k_{pq} + B_6 \, (\varepsilon_{imp} k_{jpn} +
\varepsilon_{jnp} k_{ipm}) \Big] ,
\label{KB14}
\end{eqnarray}
\end{widetext}
\noindent
$\hat k_i = k_i / k$, $\, \chi(\psi) = 1 / (1
+ \psi^2) $, $\, \psi = 2 \tau(k) \, ({\bm k}
\cdot {\bm \Omega}) / k $, $\, B_1 = 1 + \chi(2
\psi) ,$ $\, B_2 = B_1 + 2 - 4 \chi(\psi) ,$ $\,
B_3 = 2 \psi \, \chi(2 \psi) ,$ $\, B_4 = 2
\chi(\psi) - B_1 ,$ $\, B_5 = 2 - B_1 $ and $B_6
= 2 \psi \, [\chi(\psi) - \chi(2 \psi)]$, see \cite{EGKR05}.

We use the following model of the background  homogeneous stratified
turbulence or turbulent convection which takes into account an
increase of the anisotropy of turbulence with
increase of the rate of rotation:
\begin{widetext}
\begin{eqnarray}
&& f_{ij}^{(0)} \equiv \langle v_i({\bm k}_1) \,
v_j({\bm k}_2)  \rangle = {E(k)
\, [1 + 2 k \, \varepsilon_u \, \delta(\hat{\bm k}
\cdot \hat {\bm \Omega})]\over
8 \pi \, k^2 \, (k^2 + \tilde\lambda^2) \, (1 +
\varepsilon_u)}
\, \Big[\delta_{ij} \, (k^2 +
\tilde\lambda^2)  - k_i \, k_j - \tilde\lambda_i
\, \tilde\lambda_j + i \, \big(\tilde\lambda_i \,
k_j - \tilde\lambda_j \, k_i\big) \Big] \langle {\bm v}^2 \rangle,
\label{KB15}\\
&& F_{i}^{(0)} \equiv \langle v_i({\bm k}_1) \,
s({\bm k}_2)  \rangle = {3 \, E(k) \, [1 +
k \, \varepsilon_{_{F}} \, \delta(\hat{\bm k}
\cdot \hat {\bm \Omega})] \over 8 \pi \,
k^2 \, (k^2 + \tilde\lambda^2)}
\, \Big[k^2 \, e_j \, P_{ij}({\bm
k}) + i \tilde\lambda  \, k_j \, P_{ij}({\bm e})
\Big] F_\ast ,
\label{KB16}
\end{eqnarray}
\end{widetext}
\noindent
(see \cite{RK18}), and
$\Theta^{(0)} \equiv \langle s({\bm k}_1) \,
s({\bm k}_2) \rangle  =  \Theta_\ast \, E(k) / 4
\pi k^{2}$, where $F_\ast = \rho_0 \, \langle u'_z  \, s' \rangle$,
$\,  \Theta_\ast =\rho_0 \, \langle
(s')^2 \rangle$, $\delta_{ij}$ is the Kronecker tensor,
$P_{ij}({\bm e})= \delta_{ij} - e_i e_j$,
$\delta(x)$ is the Dirac delta function, $\hat{\bm k}={\bm k}/k$ and $\hat {\bm \Omega}=
{\bm \Omega}/\Omega$.
Here we have taken into account that in the anelastic approximation the velocity
fluctuations ${\bm v}= \sqrt{\rho_0}
\, {\bm u}'$ satisfy the equation $\bec{\nabla} \cdot {\bm
v} = {\bm v} \cdot \bec{\tilde\lambda}$, where
$\bec{\tilde\lambda} \equiv \bec{\lambda} / 2= -(\bec{\nabla} \rho_0) / 2\rho_0$.
To derive Eqs.~(\ref{KB15})--(\ref{KB16}) we use the following conditions:
(i) the anelastic approximation in the Fourier space implies that $(ik_i^{(1)} - \tilde\lambda_i) f_{ij}^{(0)}({\bm k},{\bm K}) = 0$, $(ik_j^{(2)} - \tilde\lambda_j) f_{ij}^{(0)} ({\bm k},{\bm K})= 0$
and $(ik_i^{(1)} - \tilde\lambda_i) F_{i}^{(0)}({\bm k},{\bm K}) = 0$,
where ${\bm k}_1 \equiv {\bm k}^{(1)} = {\bm k} + {\bm K}/2$ and ${\bm k}_2 \equiv {\bm k}^{(2)} = - {\bm k} + {\bm K}/2$;
(ii) $\int f_{ii}^{(0)} ({\bm k},{\bm K}) \exp \left[i {\bm K} \cdot {\bm R}\right] \, d {\bm k} \, d {\bm K} = \rho_0 \, \langle {\bm u}^2 \rangle^{(0)}$;
(iii)  $f_{ij}^{(0)} ({\bm k},{\bm K}) = f_{ji}^{*(0)} ({\bm k},{\bm K}) = f_{ji}^{(0)} (-{\bm k},{\bm K})$.

Solution of Eq.~(\ref{KB8}) for fast rotation by iterations
in small parameter $\ell_0 \lambda$
reads:
\begin{eqnarray}
&& \hat F^{(1,U)} = \tau^2 \left(\hat J^{U} \hat D^\lambda + \hat D^\lambda
\hat J^{U} \right)\hat F^{(0)} ,
\label{I12}
\end{eqnarray}
and
\\
\begin{widetext}
\begin{eqnarray}
&& \hat F^{(2,U)} = \tau^2 \left(\hat J^{U} \hat D^\lambda + \hat D^\lambda
\hat J^{U} \right)\hat F^{(0,\lambda)}
+ \tau^2 \biggl(\hat J^{U} \hat D^{\lambda^2}
+ \hat D^{\lambda^2} \hat J^{U} \biggr)\hat F^{(0)}
+ \tau^3 \biggl(\hat J^{U} \hat D^\lambda \hat D^\lambda + \hat D^\lambda
\hat J^{U} \hat D^\lambda
+ \hat D^\lambda \hat D^\lambda \hat J^{U}
\biggr)\hat F^{(0)} .
\nonumber\\
\label{I14}
\end{eqnarray}
\end{widetext}
\noindent
Here the contribution $\hat F^{(1,U)}$ is
linear in the ratio $\ell_0/H_\rho$ (i.e., it is linear in the parameter
$\ell_0 \lambda$), while the contributions $\hat F^{(2,U)}$ is
quadratic in $\ell_0/H_\rho$, where $H_\rho = \lambda^{-1}$,
$\hat J^U \equiv J_{ij}^U({\bm k})$, $\hat D^\lambda \equiv D_{ij}^\lambda$,
$\hat D^{\lambda^2} \equiv D_{ij}^{\lambda^2}$, the vector $\hat F^{(0)}$
is the part of $F_{i}^{(0)}$ that is a zero order in $\lambda$ [i.e., it is proportional to
$k^2 \, e_j \, P_{ij}({\bm k})$], while
the operator $\hat F^{(0,\lambda)}$ is the part of $F_{i}^{(0)}$ that is
linear in $\lambda$ [i.e., it is proportional to
$i \tilde\lambda  \, k_j \, P_{ij}({\bm e})$].
Solution of Eq.~(\ref{KB7}) for fast rotation by iterations
in small parameter $\ell_0 \lambda$
up to the second-order in this parameter is given by:
\begin{eqnarray}
&& \hat f^{(1,F)} = \tau g  \left(\hat {\bm e} \hat P \hat F^{(1,U)}
+ \hat I^{U} \tau^2 \hat {\bm e}  \hat P \hat D^\lambda \hat F^{(0)}\right),
\label{I16}\\
&& \hat f^{(1,u)} = \tau \left(\hat I^{U} \tau \hat L^\lambda + \hat L^\lambda \tau
\hat I^{U}  \right)\hat f^{(0)},
\label{I15}
\end{eqnarray}
and
\begin{widetext}
\begin{eqnarray}
\hat f^{(2,F)} &=& \tau g  \hat {\bm e} \left(\hat P \hat F^{(2,U)}
+ \hat k^\lambda \hat F^{(1,U)}\right) + \tau \hat L^\lambda \biggl(\hat f^{(1,F)}
+ \tau \hat I^{U} \tau \hat D^\lambda \hat F^{(0)}\biggr)
+ \tau g \hat I^{U} \tau^2 \hat {\bm e} \hat P \biggl[\biggl(\hat D^{\lambda^2}
+ \hat k^\lambda \hat D^\lambda\biggr)\hat F^{(0)}
+ \hat D^\lambda \hat F^{(0,\lambda)}\biggr]
\nonumber\\
&=& \tau g  \hat I^{U} \tau^2 \hat {\bm e}  \hat P \left[\left(
\hat k^\lambda \hat D^\lambda + \hat D^{\lambda^2} \right) \hat F^{(0)}
+ \hat D^\lambda \hat F^{(0,\lambda)} \right]
+ \tau^3 g \left[\hat L^\lambda \hat {\bm e} \left(\hat k^\lambda \hat J^{U} \hat F^{(0)}
+ \hat P \hat J^{U} \hat F^{(0,\lambda)} \right) + \hat L^{\lambda^2}
\hat {\bm e}  \hat P \hat J^{U} \hat F^{(0)}\right]
\nonumber\\
&& + \tau^3 g \hat {\bm e} \biggl\{ \hat P \biggl[
\left(\hat J^{U} \hat D^\lambda + \hat D^\lambda \hat J^{U} \right) \hat F^{(0,\lambda)}
+ \biggl(\hat J^{U} \hat D^{\lambda^2}
+ \hat D^{\lambda^2} \hat J^{U} \biggr)\hat F^{(0)} \biggr]
+ \hat k^\lambda \left(\hat J^{U}
\hat D^\lambda + \hat D^\lambda \hat J^{U} \right) \hat F^{(0)} \biggr\},
\label{I10}\\
\nonumber\\
\hat f^{(2,u)} &=&  \tau \left(\hat I^{U} \tau \hat L^\lambda + \hat L^\lambda \tau
\hat I^{U}  \right)\hat f^{(0,\lambda)}
+ \tau \biggl(\hat I^{U} \tau \hat L^{\lambda^2}
+ \hat L^{\lambda^2} \tau
\hat I^{U}  \biggr) \hat f^{(0)} + \tau^2 \hat L^\lambda \left(\hat I^{U} \tau \hat L^\lambda
+ \hat L^\lambda \tau \hat I^{U}\right) \hat f^{(0)} .
\label{I11}
\end{eqnarray}
\end{widetext}
\noindent
Here the contributions $\hat f^{(1,F)}$ and $\hat f^{(1,u)}$ are
linear in the ratio $\ell_0/H_\rho$, while the contributions $\hat f^{(2,F)}$ and $\hat f^{(2,u)}$ are
quadratic in $\ell_0/H_\rho$, and
$\hat {\bm e}\equiv e_i$, $\hat I^U \equiv I_{ijmn}^U$, $\hat P \equiv P_{ij}({\bm k})$,
$\hat k^\lambda \equiv k_{ij}^\lambda$, $\hat L^\lambda \equiv L_{ijmn}^\lambda$,
$\hat L^{\lambda^2} \equiv L_{ijmn}^{\lambda^2}$, and the tensor $\hat f^{(0)}$
is the part of $f_{ij}^{(0)}$ that is a zero order in $\lambda$
[i.e., it is proportional to $k^2 \, P_{ij}({\bm k})$], while
the tensor $\hat f^{(0,\lambda)}$ is the part of $f_{ij}^{(0)}$ that is
linear in $\lambda$ [i.e., it is proportional to
$i \, \big(\tilde\lambda_i \, k_j - \tilde\lambda_j \, k_i\big)$].

After integration in ${\bm k}$ space in Eqs.~(\ref{I16})--(\ref{I11})
we obtain the rotational contributions to the Reynolds stresses,
$f_{ij}=f_{ij}^{(F,\Omega)}+f_{ij}^{(u,\Omega)}$,
for the fast rotating stratified anisotropic homogeneous
turbulence or density stratified turbulent convection for the fast rotation,
where $f_{ij}^{(F,\Omega)}$ and $f_{ij}^{(u,\Omega)}$:
\begin{widetext}
\begin{eqnarray}
f_{ij}^{(F,\Omega)} &=& - A_F \, \rho_0 \, \nu_{_{T}} \, \Omega \tau_0 \,
{\ell_0^2 \over H_\rho^2} \, \biggl\{e_i e_j \meanW_z
+ 2\big(\meanW_i e_j + \meanW_j e_i\big)
+ 6\left[\left({\bm e} {\bm \times} {\bm \nabla}\right)_i e_j +
\left({\bm e} {\bm \times} {\bm \nabla}\right)_j e_i \right]
\, \meanU_z + \left({\bm e} {\bm \times} {\bm \nabla}\right)_i \meanU_j^{\perp}
\nonumber\\
&& \quad + \left({\bm e} {\bm \times} {\bm \nabla}\right)_j \meanU_i^{\perp}
+ 2\left[\nabla_i^{\perp} \left({\bm e} {\bm \times} \meanUU\right)_j
+\nabla_j^{\perp} \left({\bm e} {\bm \times} \meanUU\right)_i\right]
-4 \nabla_z\left[\left({\bm e} {\bm \times} \meanUU\right)_i e_j
+\left({\bm e} {\bm \times} \meanUU\right)_j e_i\right] \biggr\}   ,
\label{I1}\\
f_{ij}^{(u,\Omega)} &=& - {A_u \over 2} \, \rho_0 \, \nu_{_{T}} \, \Omega \tau_0 \,
{\ell_0^2 \over H_\rho^2} \, \biggl\{4\big(\meanW_i e_j
+ \meanW_j e_i\big)
+ 4\left[\left({\bm e} {\bm \times} {\bm \nabla}\right)_i e_j +
\left({\bm e} {\bm \times} {\bm \nabla}\right)_j e_i \right]
\, \meanU_z
\nonumber\\
&& \quad + 3(q+1) \left[\left({\bm e} {\bm \times} {\bm \nabla}\right)_i \meanU_j^{\perp}
+ \left({\bm e} {\bm \times} {\bm \nabla}\right)_j \meanU_i^{\perp}\right]
+ (3q+7) \left[\nabla_i^{\perp} \left({\bm e} {\bm \times} \meanUU\right)_j
+\nabla_j^{\perp} \left({\bm e} {\bm \times} \meanUU\right)_i\right]
 \biggr\}  .
\label{I2}
\end{eqnarray}
\end{widetext}
\noindent
Note that the contributions $\hat f^{(1,F)}$ and $\hat f^{(1,u)}$ (which are
linear in $\ell_0/H_\rho$) to $f_{ij}^{(F,\Omega)}$ and $f_{ij}^{(u,\Omega)}$
vanish. This implies that only the quadratic contributions, $\hat f^{(2,F)}$ and $\hat f^{(2,u)}$, in $\ell_0/H_\rho$ are the leading-order contributions to $f_{ij}^{(F,\Omega)}$ and $f_{ij}^{(u,\Omega)}$.

To integrate over the angles in ${\bm k}$-space, we use
the following integrals:
\begin{eqnarray}
&& \int k_{ij}^{\perp} \,d\varphi = \pi \delta_{ij}^{(2)},
\quad \int k_{ijmn}^{\perp} \,d\varphi = {\pi \over 4} \Delta_{ijmn}^{(2)},
\label{EC14}\\
&& \int k_{ijmnpq}^{\perp} \,d\varphi = {\pi \over 24} \Delta_{ijmnpq}^{(2)},
\label{EC15}
\end{eqnarray}
where $\delta_{ij}^{(2)}\equiv P_{ij}(\Omega) = \delta_{ij}
- \Omega_i \Omega_j /\Omega^2$,
$\Delta_{ijmn}^{(2)} = \delta_{ij}^{(2)}\delta_{mn}^{(2)}
+ \delta_{im}^{(2)} \delta_{jn}^{(2)}+ \delta_{in}^{(2)} \delta_{jm}^{(2)}$,
and
\begin{widetext}
\begin{eqnarray}
\Delta_{ijmnpq}^{(2)} = \Delta_{mnpq}^{(2)}\delta_{ij}^{(2)} +
\Delta_{jmnp}^{(2)}\delta_{iq}^{(2)} + \Delta_{imnp}^{(2)}\delta_{jq}^{(2)}
+ \Delta_{jmnq}^{(2)}\delta_{ip}^{(2)} +
\Delta_{imnq}^{(2)}\delta_{jp}^{(2)} + \Delta_{ijmn}^{(2)}\delta_{pq}^{(2)}
- \Delta_{ijpq}^{(2)}\delta_{mn}^{(2)} .
\label{EC16}
\end{eqnarray}
\end{widetext}
\noindent
Here ${\bm k}^{\perp}={\bm k} - {\bm k} \cdot \hat{\bm \Omega}$ is the wave vector in the plane perpendicular to the angular velocity ${\bm \Omega}$ with the polar angle $\varphi$ in this plane and the corresponding unit vector $\hat{\bm k}^{\perp}={\bm k}^{\perp}/k^{\perp}$. Thus, the following symmetric tensors are
defined as $k_{ij}^{\perp}=\hat k_{i}^{\perp} \hat k_{j}^{\perp}$,
$k_{ijmn}^{\perp}=k_{ij}^{\perp} \, k_{mn}^{\perp}$ and
$k_{ijmnpq}^{\perp}=k_{ij}^{\perp} \, k_{mn}^{\perp} \, k_{pq}^{\perp}$.

\section{Equations describing the large-scale instability}

To solve system of Eqs.~(\ref{C1}) and~(\ref{C2}),
we introduce new variables $\meanV(t,x,z)$ and $\meanPhi(t,x,z)$:
\begin{eqnarray}
\rho_0 \meanUU = [\meanV(t,x,z) \rho_0^{1/2}] {\bm e}_y + \bec{\nabla} {\bm
\times} \, [\meanPhi(t,x,z) \rho_0^{1/2}] {\bm e}_y,
\nonumber\\
\label{C3}
\end{eqnarray}
which corresponds to axi-symmetric problem. In the new variables
Eqs.~(\ref{C1}) and~(\ref{C2}) read
\begin{widetext}
\begin{eqnarray}
&& \left[{\partial \over \partial t} - \nu_{_{T}} \left(\Delta - {1 \over 4H_\rho^2}\right)
\right] \meanV = 2\Omega \, \biggl[\nabla_z - {1 \over 2H_\rho}
- \nu_{_{T}} \, \tau_0 \,
{\ell_0^2 \over H_\rho^3} \, \left[2 A_F \nabla_x^2 + (A_F-A_u) \left(\nabla_z^2
- {1 \over 4H_\rho^2}\right)\right]\biggr] \meanPhi,
\label{I7} \\
&& \left(\Delta - {1 \over 4H_\rho^2}\right)\, \left[{\partial \over \partial t} -
\nu_{_{T}} \left(\Delta - {1 \over 4H_\rho^2}\right) \right] \meanPhi =
- \Omega \, \biggl[2 \biggl(\nabla_z
+ {1 \over 2H_\rho}\biggr)+ \nu_{_{T}} \, \tau_0 \,
{\ell_0^2 \over H_\rho^3} \,
\biggl[(5A_F+4A_u)\nabla_x^2
\nonumber\\
&& \quad \quad \quad \quad - 2(A_F-A_u) \biggl(\nabla_z + {1 \over 2H_\rho}\biggr)^2\biggr] \biggr]
\meanV .
\label{I8}
\end{eqnarray}
\end{widetext}
\noindent
These equations allow us to study the large-scale instability
which results in generation of the mean vorticity
for different modes (see Sect.~III).

\section{The role of the centrifugal force in production of large-scale vorticity for vanishing gravity}

In this section we study production of large-scale vorticity by fast rotating
homogeneous anisotropic turbulence for vanishing gravity.
An ensemble averaging of the momentum equation yields
the equation for the mean velocity field, $\meanUU(t,{\bm x})$,
in the reference frame rotating with the constant angular velocity ${\bm \Omega}$:
\begin{eqnarray}
 {\partial \meanU_i \over \partial t} &+& (\meanUU \cdot
\bec{\nabla}) \meanU_i = - {\nabla_i \meanP \over \meanrho} + \Omega^2 r_i
\nonumber\\
&+& 2 (\meanUU \times {\bm \Omega})_i - {1 \over \meanrho} \nabla_j \overline{\rho \, u'_{i} \, u'_{j}},
\label{SA5}
\end{eqnarray}
Here $\meanP$ is the mean fluid pressure, ${\bm u}'$ are fluctuations of fluid velocity,
$\meanrho$ is the mean fluid density that satisfies the continuity equation
written in the anelastic approximation, ${\rm div} \, (\meanrho \, \meanUU) = 0$,
and the vector ${\bm r}$ is perpendicular to ${\bm \Omega}$.
The basic equilibrium is determined by $\meanUU_0=0$ and
$(\bec{\nabla} \meanP_0) / \meanrho_0 = \Omega^2 {\bm r}$ for fast rotation,
where the equilibrium fluid pressure $\meanP_0$ and density $\meanrho_0$
are related by the isothermal equation of state $\meanP_0=c_{\rm s}^2 \meanrho_0$
with a constant sound speed $c_{\rm s}$.
We use the cylindrical coordinates $(r, \varphi, z)$, where the angular
velocity ${\bm \Omega}$ is directed along the $z$ axis.
The equilibrium profile of the fluid density is given by:
\begin{eqnarray}
\meanrho_0(r) = \rho_\ast \exp \left({r^2 \over L_\Omega^2}\right) ,
\label{SA3}
\end{eqnarray}
where $L_\Omega=\sqrt{2} c_{\rm s} /\Omega$.
The second term,  $\Omega^2 r_i$, in the right hand side of Eq.~(\ref{SA5})
for the mean fluid velocity is the centrifugal force, which causes
the inhomogeneous density distribution~(\ref{SA3}) in the plane perpendicular to
the angular velocity ${\bm \Omega}$.
In the previous sections, we consider a fast rotating turbulent convection,
where in the momentum equation we have taken into the Coriolis force, but neglected
the centrifugal force. The centrifugal force should be taken
into account only when $\Omega \geq (g/R)^{1/2}$,
where $R$ is the radius (or a typical horizontal scale of the motions).

To obtain the rotational contribution to the Reynolds stress,
we use the same approach which has been applied
in previous sections, but for isothermal turbulence
(i.e., in the absence of the heat flux ${\bm F}$)
and with the inhomogeneous fluid density in
radial direction (perpendicular to ${\bm \Omega}$).
The equation for the Reynolds stress in the ${\bm k}$
space coincides with Eq.~(\ref{KB7}) in Appendix~A
with the vanishing term $\tau \,  \tilde M_{mn}^F$.
Integrating in ${\bm k}$ space in this equation, we obtain
the contribution to the Reynolds stress caused by a fast rotation:
\begin{eqnarray}
f_{ij}^{\Omega} &=&
\left[\left({\bm \Omega} {\bm \times} {\bm \lambda}^{(\Omega)}\right)_i \lambda_j^{(\Omega)} +
\left({\bm \Omega} {\bm \times} {\bm \lambda}^{(\Omega)}\right)_j \lambda_i^{(\Omega)} \right]
\, {\meanrho_0 \, u_0 \, \ell_0^3 \, \varepsilon_u \over 5(1+ \varepsilon_u)}.
\nonumber\\
\label{SB2}
\end{eqnarray}
This equation has been derived in \cite{RK18}
(see the first two terms in the right hand side of Eq.~(B13) in Appendix~B of Ref.~\cite{RK18},
where now the unit vector ${\bm e}$ is perpendicular to ${\bm \Omega}$).
Here ${\bm \lambda}^{(\Omega)} = -(\bec{\nabla} \meanrho_0) / \meanrho_0
= - {\bm r} /L_\Omega^2$ [see Eq.~(\ref{SA3})]
and the radius-vector ${\bm r}$ is perpendicular to ${\bm \Omega}$.

In the cylindrical coordinates $(r, \varphi, z)$,
the $\varphi$-component of the mean velocity is
determined by the following equation:
\begin{eqnarray}
\meanrho_0 \, {\partial \meanU_\varphi \over \partial t} &=& - {1 \over r^2}
{\partial \over \partial r} \Big[r^2 \left(f_{r \varphi}^{\Omega} + f_{r \varphi}^{\nu} \right)\Big]
+ 2 \, \meanrho_0 \, (\meanUU {\bm \times} {\bm \Omega})_\varphi ,
\nonumber\\
\label{SC1}
\end{eqnarray}
where the contribution to the Reynolds stress
caused by uniform rotation is given by
\begin{eqnarray}
f_{r \varphi}^{\Omega} = \meanrho_0 \nu_{_{T}} \, {3 \Omega \, \varepsilon_u \over 5(1+ \varepsilon_u)} \, \left({\ell_0^2 \over L_\Omega^4}\right) \,  r^2 ,
\label{SC2}
\end{eqnarray}
while the contribution to the Reynolds stress
caused by turbulent viscosity is
\begin{eqnarray}
f_{r \varphi}^{\nu} = \meanrho_0 \nu_{_{T}} r {\partial \over \partial r}
\left({\meanU_\varphi \over r} \right) .
\label{SC3}
\end{eqnarray}
For simplicity we have considered the case when the radial dependence of the mean velocity
is the strongest one, i.e., $\meanU_\varphi=\meanU_\varphi(t, r)$.
This implies that the last term in the right hand side of Eq.~(\ref{SC1}) vanishes.
We also neglect here a small kinematic viscosity in comparison with
the turbulent viscosity.

The steady-state solution of Eq.~(\ref{SC1}) reads
\begin{eqnarray}
\meanU_\varphi^{\rm (steady)}(r) = {3 \Omega \, \varepsilon_u \over 10(1+ \varepsilon_u)} \, \left({\ell_0^2 \over L_\Omega^4}\right) \, r^3 ,
\label{SC4}
\end{eqnarray}
which yields the vertical mean vorticity as
\begin{eqnarray}
\meanW_z^{\rm (steady)}(r) &\equiv& {1 \over r} {\partial \over \partial r}
\left(r \, \meanU_\varphi^{\rm (steady)}\right)
\nonumber\\
&=& {6 \Omega \, \varepsilon_u \over 5(1+ \varepsilon_u)} \, \left({\ell_0^2 \over L_\Omega^4}\right) \, r^2 .
\label{SC5}
\end{eqnarray}
Therefore, the balance between the contributions $f_{r \varphi}^{\Omega}$ to the Reynolds stress caused by a fast rotation and that caused by the turbulent viscosity, $f_{r \varphi}^{\nu}$,
determines the produced time-independent large-scale vorticity, $\meanW_z^{\rm (steady)}(r)$, see Eq.~(\ref{SC5}).

In the absence of the contribution $f_{r \varphi}^{\Omega}$ to the Reynolds
stress caused by a fast uniform rotation, Eq.~(\ref{SC1}) for
$\meanOmega_\varphi(t, r) \equiv \meanU_\varphi / r$ reads:
\begin{eqnarray}
{\partial \meanOmega_\varphi \over \partial t} &=& \nu_{_{T}} \,
\left[{\partial^2 \meanOmega_\varphi \over \partial r^2} + {3 \over r} \left(1 + {2 \, r^2 \over 3 L_\Omega^2} \right) {\partial \meanOmega_\varphi \over \partial r}\right] .
\label{SC6}
\end{eqnarray}
This equation has a decaying solution for $\meanOmega_\varphi$ caused by the turbulent viscosity:
\begin{eqnarray}
\meanOmega_\varphi (t,r) = 2 C_\ast \, \Omega \, \exp (- \gamma_{\rm dec} t) \, \Phi\left({\gamma_{\rm dec} L_\Omega^2 \over 2 \nu_{_{T}}}, 2 , -{r^2 \over 2 L_\Omega^2}\right) ,
\nonumber\\
\label{SC7}
\end{eqnarray}
where $\Phi(a, b, z)$ is the degenerate hypergeometric function, $\gamma_{\rm dec}$ is the damping rate due to the turbulent viscosity and $C_\ast$ is a free constant. For $r \ll L_\Omega$, this solution reads:
\begin{eqnarray}
\meanOmega_\varphi(t,r) = 2 C_\ast \, \Omega \, \exp (- \gamma_{\rm dec} t) \left(1 - {\gamma_{\rm dec} r^2 \over 4 \nu_{_{T}}}\right) .
\label{SC8}
\end{eqnarray}
In Eq.~(\ref{SC8}) we have to exclude a uniform rotation, so that
the vertical mean vorticity corresponding to the decaying solution is given by:
\begin{eqnarray}
\meanW_z^{\rm (decay)}(t, r) &\equiv& {1 \over r} {\partial \over \partial r}
\left(r^2 \, \meanOmega_\varphi\right)
\nonumber\\
&=& - C_\ast \, \Omega \, {\gamma_{\rm dec} r^2 \over 4 \nu_{_{T}}} \, \exp (- \gamma_{\rm dec} t) .
\label{SC9}
\end{eqnarray}
The total mean vertical vorticity, $\meanW_z^{\rm (tot)}$, is determined by the sum of homogeneous and inhomogeneous solutions of Eq.~(\ref{SC1}), i.e., $\meanW_z^{\rm (tot)}$ is given by
the sum of the stationary and decaying solutions,
$\meanW_z^{\rm (tot)} \equiv \meanW_z^{\rm (steady)} + \meanW_z^{\rm (decay)}$.
The free constant $C_\ast$ is determined by the initial condition: $\meanW_z^{\rm (tot)}(t=0) = 0$, so that:
\begin{eqnarray}
C_\ast = {6 \varepsilon_u \over 5(1+ \varepsilon_u)} \, \left({\ell_0^2 \, \nu_{_{T}} \over L_\Omega^4
\gamma}\right) .
\label{SC10}
\end{eqnarray}
Therefore, the total mean vertical vorticity for $r \ll L_\Omega$ reads:
\begin{eqnarray}
\meanW_z^{\rm (tot)} = {6 \, \Omega \, \varepsilon_u \over 5(1+ \varepsilon_u)} \, \left({\ell_0^2 \, \, r^2 \over L^{4}_{\Omega}}\right) \, \Big[1 - \exp (- \gamma_{\rm dec} t)\Big].
\nonumber\\
\label{SC12}
\end{eqnarray}
At small times, $\gamma_{\rm dec} t \ll 1$, we obtain a linear in time growing solution for the total mean vertical vorticity:
\begin{eqnarray}
\meanW_z^{\rm (tot)} = {6 \, \Omega \, \varepsilon_u \over 5(1+ \varepsilon_u)} \, \left({\ell_0^2 \, \, r^2 \over L_\Omega}\right) \, \gamma_{\rm dec} t .
\label{SC14}
\end{eqnarray}
Therefore, a combined effect of a fast rotation and horizontal inhomogeneity of the fluid density
(caused by the centrifugal force)
results in the production of the large-scale vertical vorticity
in an anisotropic turbulence. A balance between the effective force caused by
the rotational contributions to the Reynolds stress and that due to the turbulent viscosity
determines the  vertical component of the large-scale vorticity given by Eq.~(\ref{SC5}).

\end{document}